\documentclass[reqno]{amsart}
\usepackage{graphicx,epsfig}
\usepackage{epstopdf}
\usepackage{enumitem}
\usepackage{todonotes}
\usepackage[numbers,sort&compress]{natbib}
\usepackage{hyperref}
\usepackage{soul}
\usepackage{amsaddr}
\usepackage[final,color,notref,notcite]{showkeys}

\usepackage[charter]{mathdesign}

\def\d{\delta}
\def\D{\Delta}

{\rm }
\def\la{\lambda}

\def\s{\sigma}

\def\be{\begin{equation}}
 \def\ee{\end{equation}}
 \def\bea{\begin{eqnarray}}
 \def\eea{\end{eqnarray}}
 \def\bes{\begin{eqnarray}}
 \def\ees{\end{eqnarray}}
 \def\bi{\begin{itemize}}
 \def\ei{\end{itemize}}

\newcommand{\bra}[1]{\langle#1|}
\newcommand{\ket}[1]{|#1\rangle}

\renewcommand{\sec}[1]{\hyperref[sec:#1]{Sec.~\ref{sec:#1}}}

\newcommand{\fig}[1]{\hyperref[fig:#1]{Fig.~\ref{fig:#1}}}

\definecolor{labelkey}{rgb}{0,.56,.7}
\DeclareMathAlphabet{\pazocal}{OMS}{zplm}{m}{n}   

\newcommand{\Scal}{\pazocal{S}}

\newcommand{\Ncal}{\pazocal{N}}

\newcommand{\Acal}{\pazocal{A}}

\def\ve{\varepsilon}
\def\vr{\varrho}

\def\om{\omega}

\def\s{\sigma}

\def\la{\lambda}
\def\ups{\upsilon}

\newcommand*{\at}{@}
\newcommand{\dif}[1]{\mathrm{\,d} #1}             
\newcommand{\difff}[1]{\mathrm{\,d}^3 #1}             
\def\df{\overset{\mathrm{df}}{=}}
\newcommand{\Tr}[1]{\mathop{{\mathrm{Tr}}_{#1}}}
\newcommand{\diag}{\mathop{{\mathrm{diag}}}\nolimits}

\def\dg{\dagger}
\newcommand{\id}{\mathop{{\mathrm{id}}}\nolimits}
\newcommand{\kbr}[2]{| #1\rangle\!\langle #2 |}
\def\nn{\nonumber}

\DeclareSymbolFont{Eulerscripteusm10}{U}{eus}{m}{n}
\SetSymbolFont{Eulerscripteusm10}{bold}{U}{eus}{b}{n}
\DeclareMathSymbol{\euI}{\mathord}{Eulerscripteusm10}{"49}

\sodef\so{}{.065em}{.4em plus1em}{2em plus.1em minus.1em}

\begin{document}

\title[Absolutely covert quantum communication]{\so{  Absolutely covert quantum communication}}

\author{Kamil Br\'adler}
\email{kbradler\at uottawa.ca}
\address{Department of Mathematics and Statistics, University of Ottawa, Ottawa, Canada}
\author{Timjan Kalajdzievski}
\address{Department of Physics and Astronomy, York University, Toronto, Canada}
\author{George Siopsis}
\email{siopsis\at tennessee.edu}
\address{Department of Physics and Astronomy, The University of Tennessee, Knoxville, Tennessee 37996-1200, U.S.A.}
\author{Christian Weedbrook}
\address{CipherQ, 10 Dundas St E, Toronto, M5B 2G9, Canada}
\date{\today}

\begin{abstract}
We present truly ultimate limits on covert quantum communication by exploiting quantum-mechanical properties of the Minkowski vacuum in the quantum field theory framework. Our main results are the following: We show how two parties equipped with Unruh-DeWitt detectors can covertly communicate at large distances without the need of hiding in a thermal background or relying on various technological tricks. We reinstate the information-theoretic security standards for reliability of asymptotic quantum communication and show that the rate of covert communication is strictly positive. Therefore, contrary to the previous conclusions, covert and reliable quantum communication is possible.
\end{abstract}

\maketitle

\thispagestyle{empty}

\allowdisplaybreaks

\section{Introduction}

Long before the development of cryptography and encryption, one of humankind's best techniques of secretly delivering a message was to hide the fact that a message was even delivered at all. A well-known example of this is from ancient times where one person would shave another's head and write a message on it. It was only after waiting for their hair to grow back that they would then be sent to the intended recipient, who would again shave the scalp to reveal the message. This is historically one of the first-known examples of covert communication, i.e., the ability to communicate without being detected~\cite{Singh2011}.

Today, even though advanced encryption techniques are available, there is still a need to hide the fact that communication is taking place from malicious eavesdroppers, e.g., secret government operations. Modern approaches of such covert communication exist from digital steganography to spread spectrum communication and more~\cite{simon1994spread,provos2003hide,Sanguinetti2016}. Another approach consists of hiding information in the naturally occurring thermal noise of optical channels, such as additive white Gaussian noise (AWGN) channels~\cite{Che2013,Bash2013a,Bash2013,Shaw2011,Bash2015,Arrazola2016}. A natural quantum generalization of the AWGN channel is the lossy bosonic channel; a beam splitter mixing the signal with a vacuum state (pure-loss channel) or a thermal state. The lossy bosonic channel is a good model of losses for quantum-optical links used in quantum communication. Covert transmission in this case has been considered both from the point of view of a classical channel with a powerful quantum hacker~\cite{Bash2015}, as well as recently, an all-quantum channel and hacker~\cite{Arrazola2016}.
Understandably, Refs.~\cite{Bash2015,Arrazola2016} concluded that a  pure-loss channel cannot be used because  there is no way to stealthy hide the information, i.e., when the hiding medium (thermal noise) is absent. Surprisingly, the situation does not get much better if the signal is mixed with a thermal state. In Refs.~\cite{Bash2013a,Bash2015} it was shown that the asymptotic rate of covert communication is zero and, worse, the probability of a decoding error does not vanish for realistic photon detectors.

In this paper, we show that in fact no concealing medium or object is needed at all in order to hide quantum information transfer at large distances. This counterintuitive result relies on the machinery of relativistic quantum mechanics -- quantum field theory~\cite{duncan2012conceptual,birrell1984quantum}. Specifically, we show that  covert quantum communication is possible using the intrinsic resource of time-like entanglement present in Minkowski vacuum~\cite{olson2012extraction}. This is similar to the entanglement of Minkowski vacuum that can be mined at space-like distances as pioneered by~\cite{reznik2005violating} following early insights from~\cite{summers1987maximal,halvorson2000generic}. Cosmological and other consequences of space-like entanglement were investigated in~\cite{ver2009entangling,cliche2010information,svaiter1992inertial,nambu2013entanglement,jonsson2015information,salton2015acceleration} for observers equipped with an Unruh-DeWitt (UDW) detector~\cite{israel1979general} and  following various trajectories~\cite{schlicht2004considerations,louko2006often,benatti2004entanglement,higuchi1993uniformly,lin2010entanglement}. In our case the local measurements of two temporally and spatially  separated parties induce a quantum channel whose asymptotic rate of quantum communication supported by the two-way classical communication is in fact nonzero. We have calculated an upper bound on the communication rate and interpreted it as the ultimate covert communication rate. Our results offers significant differences from previous work relating to both relativistic quantum information protocols and non-relativistic quantum protocols. It was shown in~\cite{Ralph2015} (also following~\cite{olson2012extraction}) that quantum key distribution could be achieved without actually sending a quantum signal by exploiting the concept of time-like entanglement.  However, the authors of~\cite{Ralph2015} did not consider the application of covert quantum communication and also considered the use of continuous variables~\cite{Braunstein2005,Weedbrook2012} as the substrate. Our approach based on discrete variables allows us to do away with the shared coherent beam needed to synchronize the local oscillators   (needed for homodyne detection) that could  eventually blow the cover. Another problem with the oscillator mode may be its delivery especially for participants below the horizon or at large distances.
Our construction avoids this problem while reaching similarly large distances for quantum communication. For our protocol as well as for~\cite{Ralph2015} the parties have to independently solve the problem of how to covertly exchange classical messages. In general, we cannot do away with classical communication at large spatial distances but we did find instances where no classical communication is needed at all.

This paper is structured as follows. Sec.~\ref{sec:entGen} is the technical part of the paper where we introduce two inertial observers both equipped with an Unruh-DeWitt  detector. We analyze the response of the detectors to the fourth perturbative order and extract the quantum state of the two detectors after the scalar field has been traced out as the main object of our study. In Sec.~\ref{subsec:4thorderDM} we present the quantum state for zero spatial distance of the two observers subsequently generalized to a nonzero spatial distance in Sec.~\ref{subsec:4thorderLnonzero}. This is our main object of study and we interpret it from the quantum information theory point of view in Sec.~\ref{sec:keyrate}. This enables us to calculate and present our main results in Sec.~\ref{sec:results}. We conclude by mentioning open problems in Sec.~\ref{sec:discussion}.

\section{Entanglement creation via a pair of Unruh-DeWitt detectors}\label{sec:entGen}

Alice and Bob are in possession of two-state detectors whose energy levels can be made time-dependent. Tunability of energy levels can be achieved, e.g., by using a superconducting system. Thus Bob's detector obeys the Schr\"odinger equation $i\frac{\partial}{\partial t} |\Psi\rangle = \frac{H_0}{at} |\Psi\rangle$, where $H_0$ has a constant spectrum. It can be turned to a standard Schr\"odinger equation in conformal time $\tau$, defined by $t = \frac{1}{a} e^{a \tau}$. This coordinate change is equivalent to a change of the coordinate system of an inertial observer to the system of an observer undergoing uniform acceleration. Thus, Bob's detector effectively couples to Rindler modes in the future (F) wedge. Similarly, we setup Alice's detector so that it effectively couples to the Rindler modes in the past (P) wedge, by tuning the energy levels so that they correspond to the Hamiltonian $H_0$ in conformal time $\eta$, where $t = - \frac{1}{a} e^{a\eta}$. Both detectors couple to (distinct) vacuum modes which we model by a real massless scalar field $\phi$. For an introduction to such concepts see~\cite{duncan2012conceptual}.

Moreover, we are interested in the case where Alice and Bob are separated by a distance $L$. For definiteness, let $\mathbf{r} = (0,0,0)$ for Alice, and $\mathbf{r} = (L,0,0)$ for Bob. If Alice's (Bob's) detector clicks at time $t$ ($t'$), then the proper time between the two clicks is
\begin{equation}
(\Delta s)^2 = (t'-t)^2 - L^2
\end{equation}
We expect the maximal correlations in the vicinity of the null line connecting the two events. In terms of their respective conformal times, this condition can be written as $e^{a\tau} + e^{a\eta} \approx aL$. For definiteness, we choose the RHS to be $aL+2$ (similar results are obtained for any other choice). Thus, if Alice's detector performs a measurement at $\eta =0$, then optimally, Bob's detector will perform its measurement at Bob's conformal time $\tau = \tau_0$, where
\begin{equation}
\tau_0 = \frac{1}{a} \ln{[aL +1]}.
\end{equation}

The standard way of evaluating the interaction between a field and UDW detectors is perturbatively in the interaction picture. A realistic UDW detector is described by the following term
\begin{equation}\label{eq:UDWHint}
  H(t)=w(t)(\s^+e^{it\d+at}+\s^-e^{-it\d+at})\int\difff{x} f(x)\phi(t,x),
\end{equation}
where
\begin{equation}\label{eq:RealField}
  \phi(t,x)={1\over(2\pi)^3}\int{\difff{k}\over2\om_k}\big(a_ke^{-i(\om_kt-k\cdot x)}+a^\dg_ke^{i(\om_kt-k\cdot x)}\big)
\end{equation}
is a real scalar field where $(\om_k,k)$ is a four-momentum, $w(t)$ is a window function, $f(x)$ a smearing function we set to be a delta function in this work, $\s^\pm$ are the atom ladder operators and $\d$ stands for the energy gap. The evolution operator for observer $A$ reads
\begin{align}\label{eq:2ndOrderExp}
  U_A(t,t_0) & = \mathsf{T}{\Big\{\exp{\big[-i\int_{t_0}^t\dif{t'}H_A(t')\big]}\Big\}} \nn\\
  & = 1-i\int_{t_0}^t\dif{t_1}H_A(t_1)- \frac{1}{2}\mathsf{T} \bigg( \int_{t_0}^t\dif{t_1} H_A(t_1) \bigg)^2 +\frac{(-i)^3}{3!}\mathsf{T}
  \bigg( \int_{t_0}^t\dif{t_1} H_A(t_1) \bigg)^3 + \dots .
\end{align}
Considering the initial state of the field and detectors to be $\psi_{ini}=\ket{0}_A\ket{0}_B\ket{0_M}$ we calculate from $\psi_{fin}=U_A(\tau,\tau_0)U_B(\eta,\eta_0)\psi_{ini}$ (we label the time coordinate in the future/past wedge by $\tau/\eta$) the following density matrix of the two atoms after we trace over the field degrees of freedom
\begin{equation}\label{eq:DetectorMatrix}
  \om_{AB}(\kbr{x}{y})=\begin{bmatrix}
             a_1 & 0 & 0 & c_1 \\
             0 & a_2 & c_2 & 0 \\
             0 & \bar{c}_2 & b_2 & 0 \\
             \bar{c}_1 & 0 & 0 & b_1 \\
           \end{bmatrix},
\end{equation}
where the bar denotes complex conjugation. We further denote $x\in \{00,01,10,11\}_{AB}$, where $0(1)$ stands for the lower (upper) atom level. The state $\om_{AB}$ is a two-qubit density matrix. Indeed, in the Unruh-DeWitt model the atoms are two qubits as the ladder operators commute: $\s^\pm_A\s^\pm_B=\s^\pm_B\s^\pm_A$.  

Using the notation (time ordering and tensor product implied)
\begin{equation}
U_A U_B = e^{-i(A_+\sigma_A^++A_-\sigma_A^-+B_+\sigma_B^++B_-\sigma_B^-)} = \sum_{n=0}^\infty \frac{(-i)^n}{n!} (A_+\sigma_A^++A_-\sigma_A^-+B_+\sigma_B^++B_-\sigma_B^-)^n,
\end{equation}
where
\begin{align}
    A_+ & =e^{i\tau\d+a\tau}\phi(\tau,x), \\
    B_+ & =e^{-i\eta\d+a\eta}\phi(\eta,x)
\end{align}
and $A_-=A_+^\dg,B_-=B_+^\dg$. We write for the detectors' density matrix
\begin{equation}
\om_{AB}(\kbr{x}{y}) = \bra{0_M} \bra{x} U_AU_B \ket{00} \big(\bra{y} U_AU_B \ket{00}\big)^\dg \ket{0_M}.
\end{equation}
and we obtain
\begin{subequations}
\begin{align}
    \bra{00} U_AU_B \ket{00} &= 1 - \frac{1}{2} (A_+A_- + B_+B_-) + \frac{1}{24} \left( A_+^2A_-^2 + B_+^2 B_-^2 + 6 A_+A_-B_+B_- \right)+\dots\\
    \bra{01} U_AU_B \ket{00} &= -iB_+ + \frac{i}{6} \left( B_+^2 B_- + 3 A_+A_-B_+ \right) + \dots\\
    \bra{10} U_AU_B \ket{00} &= -iA_+ + \frac{i}{6} \left( A_+^2 A_- + 3 B_+B_-A_+ \right) + \dots \\
    \bra{11} U_AU_B \ket{00} &= -A_+ B_+ + \frac{1}{6} \left( A_+^2 A_- B_+ + A_+ B_+^2 B_- \right) + \dots
\end{align}
\end{subequations}
Therefore
\begin{subequations}\label{eq:upTo4thOrder}
\begin{align}
    a_1 &= 1 -  \bra{0_M} (A_+A_-+B_+B_-) \ket{0_M} + \frac{1}{4} \bra{0_M} |A_+A_-+B_+B_-|^2 \ket{0_M}\nn  \\
    & \quad+ \frac{1}{12}  \bra{0_M}\left( A_+^2A_-^2 + B_+^2 B_-^2 + 6 A_+A_-B_+B_- \right) \ket{0_M}+ \dots\\
    b_1 &=  \bra{0_M} |A_+B_+|^2\ket{0_M} + \dots \\
     c_1 &= -\bra{0_M}A_+B_+\ket{0_M} + {1\over2} \bra{0_M}A_+B_+ (A_+A_- +B_+B_-)^\dg \ket{0_M}\nn \\
     & \quad + \frac{1}{6} \bra{0_M}\left( A_+^2 A_- B_+ + A_+ B_+^2 B_- \right) \ket{0_M}+ \dots\\
     a_2 &= \bra{0_M} |B_+|^2 \Big[ 1 - \frac{1}{3}  (B_+B_-+ 3A_+A_-) \Big] \ket{0_M} + \dots\\
     b_2 &= \bra{0_M} |A_+|^2 \Big[ 1 - \frac{1}{3}  (A_+A_-+ 3B_+B_-) \Big] \ket{0_M} + \dots\\
     c_2 &= \bra{0_M} A_+^\dg B_+ \Big[ 1 - \frac{1}{6} (A_+A_-+ 3B_+B_-)^\dg - \frac{1}{6} (B_+B_- + 3A_+A_-)\Big] \ket{0_M} + \dots,
\end{align}
\end{subequations}
where e.g. $|A_+|^2$ means $A_+A_+^\dg=A_+A_-$.

\subsection{Second order density matrix expansion  for a Gaussian window function}\label{sec:window}
We will need the following standard results~\cite{birrell1984quantum}
\begin{equation}\label{eq:2pointCorr}
  \bra{0_M} \mathsf{T} \left( \phi(t_1,x_1)\phi(t_2,x_2)\right) \ket{0_M}\equiv\Delta(X_1,X_2)=-{1\over4\pi^2}{1\over(t_1-t_2-i\varepsilon)^2-(x_1-x_2)^2},
\end{equation}
where $\Delta(X_1,X_2)$ stands for the Feynman propagator and $X_i$ denotes a four vector. Following~\cite{olson2012extraction}, the FF (and $\eta$ for PP) propagator reads
\begin{equation}\label{eq:PropagFFandPP}
  \Delta^{FF}(\tau,\tau')=-{a^2\over16\pi^2}{e^{-a(\tau+\tau')}\over\sinh^2{[a(\tau-\tau')/2-i\ve]}}
\end{equation}
and the FP propagator takes the following form
\begin{equation}\label{eq:PropagFP}
  \Delta^{FP}(\tau,\eta)=-{a^2\over16\pi^2}{e^{-a(\tau+\eta)}\over\cosh^2{[a(\tau-\eta)/2-i\ve]}}.
\end{equation}
Propagators Eqs.~(\ref{eq:PropagFFandPP}) and~(\ref{eq:PropagFP}) can be further simplified using the identities
\begin{subequations}
\begin{align}
  {1\over\sinh^{2}{x\over2}} & = 4\sum_{n=-\infty}^\infty{1\over(x+2\pi i n)^2}, \\
  {1\over\cosh^{2}{x\over2}} & = -4\sum_{n=-\infty}^\infty{1\over(x+\pi i (2n+1))^2}.
\end{align}
\end{subequations}
We obtain
\begin{equation}\label{eq:PropagFFandPP1}
  \Delta^{FF}(\tau,\tau')=-{e^{-a(\tau+\tau')}\over4\pi^2}\sum_{n=-\infty}^\infty{1\over(\tau-\tau'-2i\ve+2\pi in/a)^2}
\end{equation}
and (note that we can set $\ve=0$ here right away)
\begin{equation}\label{eq:PropagFP1}
  \Delta^{FP}(\tau,\eta)={e^{-a(\tau+\eta)}\over4\pi^2}\sum_{n=-\infty}^\infty{1\over(\tau-\eta-2i\ve+\pi i/a (2n+1))^2}.
\end{equation}
The matrix entries needed for up to the second order read
\begin{subequations}\label{eq:MtrxElmns2ndOrder}
\begin{align}
  \euI_1&=\bra{0_M}A_+A_-\ket{0_M}  = \int_{-\infty}^\infty\dif{\tau}\,w(\tau)\int_{-\infty}^\infty\dif{\tau'}\,w(\tau')e^{a(\tau+\tau')}e^{i\d(\tau'-\tau)}
  \bra{0_M}\mathsf{T} \{\phi(\tau,x_1)\phi(\tau',x_2)\}\ket{0_M},
  \label{eq:MtrxElmns2ndOrderApAm}\\
  \euI_2&=\bra{0_M}A_+B_+\ket{0_M} = \int_{-\infty}^\infty\dif{\eta}\,w(\eta)\int_{-\infty}^\infty\dif{\tau}\,w(\tau)e^{a(\eta+\tau)}e^{i\d(-\eta+\tau)}\bra{0_M}\mathsf{T}\{\phi(\tau,x_1)\phi(\eta,x_2)\}\ket{0_M}, \label{eq:MtrxElmns2ndOrderApBp}\\
  \euI_3&=\bra{0_M}  A_+^\dg B_+ \ket{0_M}  = \int_{-\infty}^\infty\dif{\tau}\,w(\tau)\int_{-\infty}^\infty\dif{\eta}\,w(\eta)e^{a(\tau+\eta)}e^{-i\d(\eta+\tau)}\bra{0_M}\mathsf{T}\{\phi(\tau,x_1)\phi(\eta,x_2)\}\ket{0_M}.
  \label{eq:MtrxElmns2ndOrderApdgBp}
\end{align}
\end{subequations}
For the fourth order in the next section we will also need
\begin{equation}
  \euI_4=\bra{0_M}A_-A_-\ket{0_M}  = \int_{-\infty}^\infty\dif{\tau}\,w(\tau)\int_{-\infty}^\infty\dif{\tau'}\,w(\tau')e^{a(\tau+\tau')}e^{i\d(\tau'+\tau)}
  \bra{0_M}\mathsf{T} \{\phi(\tau,x_1)\phi(\tau',x_2)\}\ket{0_M}.
\end{equation}
To evaluate the integrals we adopt the results from~\cite{sriramkumar1996finite} where one of the investigated window functions is a Gaussian profile
\begin{equation}\label{eq:GaussianProfile}
  w(t)=\la\exp{\Big[{-{t^2\over\s^2}}\Big]},
\end{equation}
where $0\leq\la\leq1$ tunes the coupling strength. The matrix elements~(\ref{eq:MtrxElmns2ndOrder}) are thus given by integrals of the following four types:
\begin{align}\label{eq:integral14}
  \euI_{4,1}&=\la^2\int_{-\infty}^\infty\dif{\tau}\,w(\tau)\int_{-\infty}^\infty\dif{\tau'}\,w(\tau')e^{a(\tau+\tau')}e^{i\d(\tau'\pm\tau)}\bra{0_M}\mathsf{T}\{\phi(\tau,x)\phi(\tau',x)\}\ket{0_M}\nn\\
  &=-{\la^2\over4\pi^2}\int_{-\infty}^\infty\dif{\tau}\,\exp{\Big[{-{\tau^2\over\s^2}}\Big]}\int_{-\infty}^\infty\dif{\tau'}\,\exp{\Big[{-{\tau'^2\over\s^2}}\Big]}e^{i\d(\tau'\pm\tau)}
  \sum_{n=-\infty}^\infty{1\over(\tau-\tau'-2i\ve+2\pi in/a)^2}
\end{align}
and
\begin{align}\label{eq:integral23}
  \euI_{3,2}&=\la^2\int_{-\infty}^\infty\dif{\eta}\,w(\eta)\int_{-\infty}^\infty\dif{\tau}\,w(\tau)e^{a(\eta+\tau)}e^{i\d(\eta\pm\tau)}\bra{0_M}\mathsf{T}\{\phi(\tau,x)\phi(\eta,x')\}\ket{0_M}\nn\\
  &={\la^2\over4\pi^2}\int_{-\infty}^\infty\dif{\tau}\,\exp{\Big[{-{\tau^2\over\s^2}}\Big]}\int_{-\infty}^\infty\dif{\eta}\,\exp{\Big[{-{\eta^2\over\s^2}}\Big]}e^{i\d(\eta\pm\tau)}
  \sum_{n=-\infty}^\infty{1\over(\tau-\eta-2i\ve+\pi i/a (2n+1))^2}.
\end{align}
Integrals $\euI_1$ and $\euI_2$ are integral (50) in~\cite{sriramkumar1996finite} (for $a=g$). Conveniently, this integral has sort of a closed form.  
Following~\cite{sriramkumar1996finite}, the inner integral splits in two expressions according to whether the poles are positive ($+$) or not ($-$) leading to
\begin{equation}\label{eq:innerInt}
  \ups_{\pm}(n)=\pm2\pi(k-\d)\exp{\Big[{-}{2\pi n(k-\d)\over a}\Big]}.
\end{equation}
Then, we get the following expressions
\begin{align}\label{eq:Int1plus}
   \euI_{1}^+&=-\la^2{\s^2\over8\pi^2}\sum_{n=-\infty}^0\int_{-\infty}^\d\dif{k}\exp{\Big[{-{k^2\s^2\over2}}\Big]}\ups_{+}(n)\nn\\
  &={\la^2\over8\pi}e^{-{(\s\d)^2\over2}}\sum_{n=-\infty}^0\Big[2+\sqrt{2\pi}\exp{\Big[{(2n\pi/(a\s)+\s\d)^2\over2}\Big]}(2n\pi/(a\s)+\s\d)
  \Big(1+\mathrm{Erf\,}\Big[{2n\pi/(a\s)+\s\d\over\sqrt{2}}\Big]\Big)\Big]
\end{align}
and
\begin{align}\label{eq:Int1minus}
   \euI_{1}^-&=-\la^2{\s^2\over8\pi^2}\sum_{n=1}^\infty\int_{\d}^\infty\dif{k}\exp{\Big[{-}{k^2\s^2\over2}\Big]}\ups_{-}(n)\nn\\
  &=-{\la^2\over8\pi}e^{-{(\s\d)^2\over2}}\sum_{n=1}^\infty\Big[{-2}+\sqrt{2\pi}\exp{\Big[{(2n\pi/(a\s)+\s\d)^2\over2}\Big]}(2n\pi/(a\s)+\s\d)
  \Big(1-\mathrm{Erf\,}\Big[{2n\pi/(a\s)+\s\d\over\sqrt{2}}\Big]\Big)\Big].
\end{align}
Note that (\ref{eq:Int1minus}),(\ref{eq:Int1plus}) and all other matrix elements we calculate will be expressed in terms of certain dimensionless parameters such as $a\s$ and $\s\d$. To get $\euI_{2}^\pm$ we swap $2n$ for $2n+1$ in~(\ref{eq:innerInt}) but we  must also shift the sums' limits because the poles in~(\ref{eq:integral23}) are vertically shifted by one:
\begin{subequations}\label{eq:Int2plusminus}
\begin{align}
   \euI_{2}^+&=-\la^2{\s^2\over8\pi^2}{\pi\over\s^3}e^{-{(\s\d)^2\over2}}\nn\\
   &\times\sum_{n=-\infty}^{-1}\Big[2\s+\sqrt{2\pi}\exp{\Big[{((2n+1)\pi/a+\s^2\d)^2\over2\s^2}\Big]}((2n+1)\pi/a+\s^2\d)
  \Big(1+\mathrm{Erf\,}\Big[{(2n+1)\pi/a+\s^2\d\over\s\sqrt{2}}\Big]\Big)\Big]\nn\\
  &=-{\la^2\over8\pi}e^{-{(\s\d)^2\over2}}\nn\\
  &\times\sum_{n=-\infty}^{-1}\Big[2+\sqrt{2\pi}\exp{\Big[{((2n+1)\pi/(a\s)+\s\d)^2\over2}\Big]}((2n+1)\pi/(a\s)+\s\d)
  \Big(1+\mathrm{Erf\,}\Big[{(2n+1)\pi/(a\s)+\s\d\over\sqrt{2}}\Big]\Big)\Big].\\
   \euI_{2}^-&=\la^2{\s^2\over8\pi^2}{\pi\over\s^3}e^{-{(\s\d)^2\over2}}\nn\\
   &\times\sum_{n=0}^\infty\Big[{-2\s+\sqrt{2\pi}\exp{\Big[{((2n+1)\pi/a+\s^2\d)^2\over2\s^2}\Big]}((2n+1)\pi/a+\s^2\d)}
   \Big({1-\mathrm{Erf\,}\Big[{{(2n+1)\pi/a+\s^2\d\over\s\sqrt{2}}}\Big]}\Big)\Big]\nn\\
   &={\la^2\over8\pi}e^{-{(\s\d)^2\over2}}\nn\\
  &\times\sum_{n=0}^{\infty}\Big[{-2}+\sqrt{2\pi}\exp{\Big[{((2n+1)\pi/(a\s)+\s\d)^2\over2}\Big]}((2n+1)\pi/(a\s)+\s\d)
  \Big(1-\mathrm{Erf\,}\Big[{(2n+1)\pi/(a\s)+\s\d\over\sqrt{2}}\Big]\Big)\Big].
\end{align}
\end{subequations}
Next, the plus version of~Eq.~(\ref{eq:integral23}) is called $\euI_3$:
\begin{align}\label{eq:Int3}
  \euI_3
  &={\la^2\over8\pi^2}\int_{-\infty}^\infty\dif{y}\,
  e^{i\d y}\exp{\Big[{-{y^2\over2\s^2}}\Big]}\int_{-\infty}^\infty\dif{x}\exp{\Big[{-{x^2\over2\s^2}}\Big]}
  \sum_{n=-\infty}^\infty{1\over(x-2i\ve+\pi i/a (2n+1))^2}.
\end{align}
It can be written as $\euI_3=\euI_{3}^++\euI_{3}^-$ where
\begin{subequations}\label{eq:Int3plusminus}
\begin{align}
  \euI_{3}^+&={\la^2\over8\pi^2}\sqrt{2\pi}\s e^{-{(\s\d)^2\over2}}\sum_{n=0}^\infty
  \Big[{-{\sqrt{2\pi}\over\s}+{(2n+1)\pi^2\over\s^2a}\exp{\Big[{(2n+1)^2\pi^2\over2\s^2a^2}\Big]}\Big(1-\mathrm{Erf\,}\Big[{(2n+1)\pi\over\sqrt{2}a\s}\Big]\Big)}\Big]\nn\\
  &={\la^2\over8\pi}e^{-{(\s\d)^2\over2}}\sum_{n=0}^\infty
  \Big[{{-2}+{(2n+1)\pi\sqrt{2\pi}\over\s{a}}\exp{\Big[{(2n+1)^2\pi^2\over2(\s{a})^2}\Big]}\Big(1-\mathrm{Erf\,}\Big[{(2n+1)\pi\over\sqrt{2}a\s}\Big]\Big)}\Big],\\
  \euI_{3}^-&={\la^2\over8\pi} e^{-{(\s\d)^2\over2}}\sum_{n=-\infty}^{-1}
  \Big[{-2-{(2n+1)\pi\sqrt{2\pi}\over\s{a}}\exp{\Big[{(2n+1)^2\pi^2\over2(\s{a})^2}\Big]}\Big(1+\mathrm{Erf\,}\Big[{(2n+1)\pi\over\sqrt{2}a\s}\Big]\Big)}\Big]=\euI_{3}^+.
\end{align}
\end{subequations}
By taking only the zeroth and second order  contributions from Eqs.~(\ref{eq:upTo4thOrder}) the $\euI_i$ expressions reveal the density matrix up to the second perturbative order:
\begin{subequations}\label{eq:2ndOrderEntries}
\begin{align}
  a_1 & = 1-2\,\euI_{1},\\
  b_1 & = 0,\\
  c_1 & = -\euI_{2},\\
  a_2 & = b_2 = \euI_{1},\\
  c_2 & = \euI_{3}.
\end{align}
\end{subequations}
The result coincides with~\cite{reznik2005violating} and has been rederived in the timelike scenario as well~\cite{olson2012extraction}.

\subsection{Fourth order density matrix expansion}\label{subsec:4thorderDM}

The second order is insufficient for our purposes since the `outer' density matrix $\om_{AB}$ is negative unless $c_1=0$. In order to evaluate the coefficients from Eqs.~(\ref{eq:upTo4thOrder}) we
first calculate the plus version of~Eq.~(\ref{eq:integral14})
\begin{align}\label{eq:Int4}
  \euI_4 &=-{\la^2\over8\pi^2}\int_{-\infty}^\infty\dif{y}\,
  e^{i\d y}\exp{\Big[{-{y^2\over2\s^2}}\Big]}\int_{-\infty}^\infty\dif{x}\exp{\Big[{-{x^2\over2\s^2}}\Big]}
  \sum_{n=-\infty}^\infty{1\over(x-2i\ve+2n\pi i/a)^2}
\end{align}
where we split it into two parts $\euI_4=\euI_{4}^++\euI_{4}^-$ calculated as
\begin{subequations}\label{eq:Int4plusminus}
\begin{align}
  \euI_{4}^+&=-{\la^2\over8\pi^2}\sqrt{2\pi}\s e^{-{(\s\d)^2\over2}}\sum_{n=0}^\infty\Big[{-{\sqrt{2\pi}\over\s}+{2n\pi^2\over\s^2a}\exp{\Big[{2n^2\pi^2\over\s^2a^2}\Big]}\Big(1-\mathrm{Erf\,}\Big[{n\pi\sqrt{2}\over\s{a}}\Big]\Big)}\Big]\nn\\ &=-{\la^2\over8\pi}e^{-{(\d\s)^2\over2}}\sum_{n=0}^\infty\Big[{{-2}+{2n\pi\sqrt{2\pi}\over{a\s}}\exp{\Big[{2n^2\pi^2\over\s^2a^2}\Big]}\Big(1-\mathrm{Erf\,}\Big[{n\pi\sqrt{2}\over\s{a}}\Big]\Big)}\Big],\\
  \euI_{4}^-&=-{\la^2\over8\pi^2}\sqrt{2\pi}\s e^{-{(\s\d)^2\over2}}\sum_{n=-\infty}^{-1}
  \Big[{-{\sqrt{2\pi}\over\s}-{2n\pi^2\over\s^2a}\exp{\Big[{2n^2\pi^2\over\s^2a^2}\Big]}
  \Big(1+\mathrm{Erf\,}\Big[{n\pi\sqrt{2}\over\s a}\Big]\Big)}\Big]\nn\\
  &=-{\la^2\over8\pi}e^{-{(\d\s)^2\over2}}\sum_{n=-\infty}^{-1}\Big[{{-2}-{2n\pi\sqrt{2\pi}\over{a\s}}\exp{\Big[{2n^2\pi^2\over\s^2a^2}\Big]}
  \Big(1+\mathrm{Erf\,}\Big[{n\pi\sqrt{2}\over\s{a}}\Big]\Big)}
  \Big].
\end{align}
\end{subequations}
We notice at once that $\euI_{4}^-$ equals $\euI_{4}^+$ without the $n=0$ term.

To proceed we use the following four-point correlation function identity as the simplest case of the Wick theorem~\cite{duncan2012conceptual}:
\begin{align}\label{eq:4pointCorr}
\bra{0_M}\mathsf{T} \bigg\{ \prod_{i=1}^4\phi(X_i) \bigg\} \ket{0_M}=\Delta(X_1,X_2)\Delta(X_3,X_4)+\Delta(X_1,X_3)\Delta(X_2,X_4)+\Delta(X_1,X_4)\Delta(X_2,X_3).
\end{align}
As a result, from Eqs.~(\ref{eq:upTo4thOrder}) we  get
\begin{subequations}\label{eq:HigherOrderOmAB}
\begin{align}
   a_1 & = 1-2\,\euI_{1}+\euI_{1}^2+\euI_{2}^2+\euI_{3}^2+{2\over3}(2\,\euI_1^2+\euI_4^2),\\
   b_1 & = \euI_{1}^2+\euI_{2}^2+\euI_{3}^2,\\
   c_1 & = -\euI_{2}+{4\over3}(2\,\euI_1\euI_2+\euI_3\euI_4),\\
   a_2 & = b_2 = \euI_{1} -(\euI_{1}^2+\euI_{2}^2+\euI_{3}^2)-{1\over3}(2\,\euI_1^2+\euI_4^2), \\
   c_2 & = \euI_{3}-{4\over3}(2\,\euI_1\euI_3+\euI_2\euI_4).
\end{align}
\end{subequations}
Note that similarly to the second perturbative order in Eq.~(\ref{eq:2ndOrderEntries}) the trace of $\om_{AB}$ assembled from (\ref{eq:HigherOrderOmAB}) equals one without a renormalization.

\subsection{Timelike entanglement for non-zero distances}\label{subsec:4thorderLnonzero}

\begin{figure}[t]
   \resizebox{10cm}{!}{\includegraphics{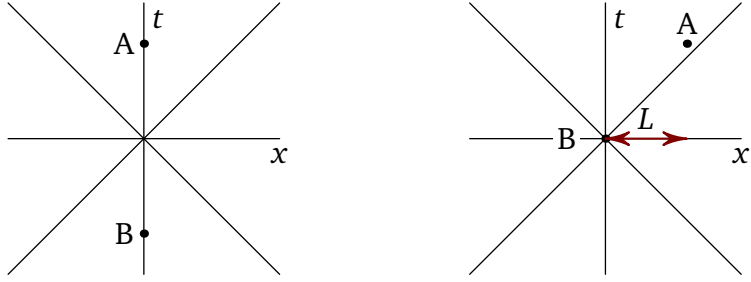}}
    \caption{Two timelike observers at the same spatial point (on the left) and separated by a spatial distance $L$ (on the right).}
    \label{fig:timelike}
\end{figure}

The timelike entanglement would be hardly useful for covert quantum communication if the stationary sender and receiver were forced to be stationed at the same spatial coordinate. Fortunately, the timelike entanglement of the scalar massless particles persists over nonzero distances. In order to quantify how much entanglement survives we take the two-point correlation function Eq.~(\ref{eq:2pointCorr}) and by fixing $x$ we have for the receiver (see Fig.~\ref{fig:timelike})
\begin{equation}
t_1={1\over a}e^{a\tau}.
\end{equation}
By setting $x_2=0,x_1=L$ propagator (\ref{eq:PropagFP}) then becomes
\begin{equation}\label{eq:PropagFPdistance}
  \D^{FP}(\tau,\eta;L)=-{a^2\over16\pi^2}{e^{-a(\tau+\eta)}\over\cosh^2{[a/2(\tau-\eta)]}-{a^2L^2\over4}\exp{[-a(\tau+\eta)]}}.
\end{equation}
We removed the $-i\ve$ prescription as no pole lies on the real axis and after rescaling $\tau\to\tau/a$ and $\eta\to\eta/a$  and plugging Eq.~(\ref{eq:PropagFPdistance}) the integrals
\begin{equation}
  \euI_{3,2}=-{\la^2\over16\pi^2}\int_{-\infty}^\infty\dif{\eta}\,e^{-{\eta^2\over(a\s)^2}}\int_{-\infty}^\infty\dif{\tau}\,e^{-{(\tau-a\tau_0)^2\over(a\s)^2}}
  e^{i{\d\over a}(\eta\pm\tau)}
  {1\over\cosh^2{[{\tau-\eta\over2}]}-{a^2L^2\over4}\exp{[-(\tau+\eta)]}}
\end{equation}
converge without the need to regularize. We have $a\tau_0=\ln{[aL+1]}$, the integrals are dimensionless and $c=1$ is used.
After  changing the variables to $\tau_\pm=\tau-a\tau_0\pm\eta$  we obtain
\begin{align}\label{eq:J23forL}
  \euI_{3,2} & = -{\la^2e^{\pm i\d\tau_0}\over32\pi^2}\int_{-\infty}^\infty\dif{\tau_+}e^{-{\tau_+^2\over2(a\s)^2}}
  \int_{-\infty}^\infty\dif{\tau_-}e^{-{\tau_-^2\over2(a\s)^2}}e^{\pm i{\d\over a}\tau_\pm}
  {1\over\cosh^2{\tau_-+a\tau_0\over2}-{a^2L^2\over4}e^{-(\tau_++a\tau_0)}}.
\end{align}

\subsection{A necessary condition for entanglement}

Looking at Eqs.~(\ref{eq:HigherOrderOmAB}) we notice an interesting fact. The off-diagonal elements $c_i$ are invariant with respect to $\euI_2\leftrightharpoons\euI_3$ (possibly up to a sign change). This means that if $\euI_2\cong\euI_3$ there can't be any entanglement -- the state $\om_{AB}$ is (nearly) PPT invariant. So in order to have entanglement (keeping in mind that the PPT non-invariance is a necessary but not a sufficient condition) we need $\euI_2\gtrless\euI_3$. We now find out when this is possible. We have
\begin{equation}
\euI_2 = \frac{\la^2}{4\pi^2}\int_{-\infty}^{\infty} \dif{\tau}\dif{\eta}\,e^{- \frac{\tau^2+\eta^2}{\sigma^2}} e^{-i\delta (\tau - \eta)} \sum_{n=-\infty}^\infty \frac{1}{(\tau -\eta + (2n+1)\pi /a)^2}.
\end{equation}
Switching to $\tau_\pm = \tau \pm\eta$ we get
\begin{equation}
\euI_2 = \frac{\la^2}{4\pi^2}\int_{-\infty}^{\infty} \dif{\tau}_+ e^{- \frac{\tau_+^2}{2\sigma^2}} \int_{-\infty}^{\infty}
\dif{\tau}_-e^{-\frac{\tau_-^2}{2\sigma^2}} e^{-i\delta \tau_- } \sum_{n=-\infty}^\infty \frac{1}{(\tau_- + (2n+1)\pi i /a)^2}.
\end{equation}
After shifting $\tau_-$, we obtain
\begin{equation}
\euI_2 = \la^2\frac{e^{-\delta^2\sigma^2/2}}{4\pi^2}\int_{-\infty}^{\infty} \dif{\tau}_+ e^{- \frac{\tau_+^2}{2\sigma^2}} \int_{-\infty}^{\infty}  \dif{\tau}_- e^{-\frac{\tau_-^2}{2\sigma^2}} \sum_{n=-\infty}^\infty \frac{1}{(\tau_- -i\delta\sigma^2+ (2n+1)\pi i /a)^2}.
\end{equation}
Turning to $\euI_3$, we have (after shifting $\tau_+$)
\begin{equation}
\euI_3 = \la^2\frac{e^{-\delta^2\sigma^2/2}}{4\pi^2}\int_{-\infty}^{\infty} \dif{\tau}_+ e^{-\frac{\tau_+^2}{2\sigma^2}}  \int_{-\infty}^{\infty}  \dif{\tau}_-e^{-\frac{\tau_-^2}{2\sigma^2}} \sum_{n=-\infty}^\infty \frac{1}{(\tau_- + (2n+1)\pi i /a)^2}.
\end{equation}
The ratio is
\begin{equation}
\frac{\euI_3}{\euI_2} =  \frac{\sum_{n=-\infty}^\infty\limits \int_{-\infty}^{\infty}\limits  \dif{\tau}_- e^{- \frac{\tau_-^2}{2\sigma^2}} \frac{1}{(\tau_- + (2n+1)\pi i /a)^2}}{\sum_{n=-\infty}^\infty\limits\int_{-\infty}^{\infty}\limits  \dif{\tau}_- e^{- \frac{\tau_-^2}{2\sigma^2}}  \frac{1}{(\tau_- -i\delta\sigma^2+ (2n+1)\pi i /a)^2}}.
\end{equation}
Each integral can be written explicitly in terms of error functions as follows. Let's set
\begin{equation}
I(\xi) = \int_{-\infty}^{\infty}  \dif{\tau}_- e^{- \frac{\tau_-^2}{2\sigma^2}}  \frac{1}{(\tau_- +i\xi)^2}
\end{equation}
and write it as a Fourier transform (omitting an overall factor):
\begin{equation}
I(\xi) \sim \int_{-\infty}^{\infty}  \dif{z} \dif{\tau}_- e^{- \frac{\sigma^2z^2}{2}} e^{iz\tau_-} \frac{1}{(\tau_- +i\xi)^2}.
\end{equation}
The integral in $\tau_-$ needs to be closed in the upper half plane if $z >0$, and in the lower half plane if $z <0$. Thus for $\xi >0$, only $z < 0$ contributes, and for $\xi <0$, only $z >0$ contributes. Therefore,
\begin{equation}
I(\xi ) \sim \int_0^{\infty} \dif{z}\, z e^{- \frac{\sigma^2z^2}{2}} e^{-|\xi| z}.
\end{equation}
We deduce (after re-scaling $z$)
\begin{equation}
\frac{\euI_3}{\euI_2} =  \frac{\sum_{n=-\infty}^\infty\limits\int_0^{\infty}\limits  \dif{z}\, z e^{- \frac{z^2}{2}} e^{-|2n+1|\pi  z/(a\sigma)}}{\sum_{n=-\infty}^\infty\limits \int_0^{\infty}\limits  \dif{z}\, z e^{- \frac{z^2}{2}} e^{-|\delta\sigma - (2n+1)\pi  /(a\sigma)|z}}.
\end{equation}
Clearly, $\euI_3 < \euI_2$. As $\sigma\to\infty$, we have $\euI_3/\euI_2 \to 0$. So for large $\sigma$, $\euI_3 \ll \euI_2$. If the symmetry between $c_1$ and $c_2$ remains even for an arbitrary perturbative order, only under this condition we may expect the creation of entanglement with the help of two UDW detectors.

\section{One- and Two-way secret key rates}\label{sec:keyrate}

The straightforward but crucial observation we will use is that the state $\om_{AB}$ can be interpreted as a Choi state representing a completely positive map (quantum channel $\Ncal_{A'\to B}$) using the Choi-Jamio\l kowski representation~\cite{choi1975completely,jamiolkowski1972linear}. Recall that given a Choi matrix $\om_\Ncal$  we find the channel action via the following isomorphism:
\begin{equation}\label{eq:JamiChoi}
  \Ncal(\vr_A)=\Tr{A}\big[(\vr_A^\top\otimes\mathrm{id}_{B})\,\om_\Ncal\big],
\end{equation}
where $A\simeq A'$ and $\top$ denotes  transposition in the canonical basis. Choi states are routinely used to evaluate the secret key rates in virtually all DVQKD protocols and as we will see it is arguably one of the best channel representations to perform various capacity calculations or estimates. This often consists of calculating the eigenvalues of the Choi matrix and its marginal states. The reducible form of $\om_\Ncal\equiv\om_{AB}$ plays into our cards: the eigenvalues of both $\om_B$ and $\om_{AB}$ are easy to calculate. The former can be directly read out from $\om_B=\diag{[a_1+b_2, a_2+b_1]}$ and for $\om_{AB}$ we get a familiar expression for all four eigenvalues:
$\la_{i}^\pm(\om_{AB})= 1/2 \big(a_i+b_i\pm\sqrt{\left(a_i-b_i\right){}^2+4 |c_i|^2}\big)$.

Let's see how useful this representation is. Given a quantum noisy channel $\Ncal$, the question of the highest achievable rate at which two legitimate parties can asymptotically establish secret correlations using one-way ($\rightarrow$) classical (public) communication was answered in~\cite{devetak2005private,rennersecurity}. It is given by the private channel capacity which is equivalent to the  one-way channel secret key rate $K_\rightarrow(\Ncal)$. The secret key rate is hard to calculate, though. Fortunately, a number of lower bounds are known. One such a bound is the coherent information $Q(\Ncal,\vr_{A'})_\om=H(B)_\om-H(E)_\om=H(B)_\om-H(AB)_\om$, where $\Ncal_{A'\to B}(\vr)=\om$ and $H(A)_\om\df-\Tr{}[\om_A\log{\om_A}]$ is the von Neumann entropy ($\log$ is to the base two throughout the paper and the symbol $\df$ stands for defined). Hence, the coherent information is  optimized over an input  state $\vr_{A'}$ to the channel $\Ncal_{A'\to B}$ but evaluated on the output states $\om_B$ and $\om_{AB}$. We are led to the following chain of inequalities~\cite{devetak2005private}:
\begin{equation}\label{eq:IneqChain1}
Q(\Ncal,\vr_{A'})_\om\leq Q^{(1)}(\Ncal)\leq Q(\Ncal)\leq K_\rightarrow(\Ncal),
\end{equation}
where $Q^{(1)}(\Ncal)\df\max_{\vr_{A'}}Q(\Ncal,\vr_{A'})$ and $Q(\Ncal)\df\lim_{n\to\infty}{1\over n}\max_{\vr_{A'^n}}Q(\Ncal^{\otimes n},\vr_{A'^n})$ are the single- and multi-letter quantum capacity formulas, respectively. In short, quantum capacity can be operationally interpreted as the  rate at which maximally entangled states can be distilled using one-way classical communication and local operations (one-way LOCC). Hence, it provides a natural lower bound on the one-way secret key rate since any quantum or classical correlation (namely a classical key -- a secret string of bits) can be  teleported once maximally entangled states are available. Unfortunately, the  regularization makes $Q(\Ncal)$ and $K_\rightarrow(\Ncal)$ practically incalculable for plenty of interesting channels.

A noteworthy peculiarity that we will encounter for other measures as well, is the inability to optimize $Q(\Ncal,\vr_{A'})$ over the channel input state $\vr_{A'}$ to get $Q^{(1)}(\Ncal)$. This is because there is no accessible input state in the first place. In quantum Shannon theory, the state $\vr_{A'}$ (or its purification $\vr_{AA'}$) is a quantum code chosen such that the entanglement generation rate (or, equivalently, quantum communication rate) is the best possible. Here we have only an image of such a state and we can  assume that the input purified state $\vr_{AA'}$ is maximally entangled. Then the output state $\om_{AB}$ is a Choi state as we interpreted it. We could take a different  point of view and decide that $\om_{AB}$ is an image of any input state in which case it is formally not a Choi state. But by this we would deprive ourselves of being able to use the secret key rate estimates introduced further in this section. There, the Choi matrix plays a crucial role. So let's adopt the following convention: In the expressions to come,c the state $\om$ or $\om_{AB}$ means a generic bipartite state. If we need to stress that we interpret $\om$ as a Choi-Jamio\l kowski matrix of a channel $\Ncal_{A'\to B}$ we will write $\om_\Ncal$ instead.

One-way classical communication is a strictly weaker resource for quantum communication than two-way ($\leftrightarrow$) classical communication. As a matter of fact, the problem of calculating the two-way secret key rate (or quantum/private capacity) recently received a lot of deserved attention~\cite{pirandola2015ultimate,takeoka2014squashed,christandl2016relative} drawing on early results from~\cite{bennett1996mixed,christandl2006quantum}. The two-way quantities of a general quantum channel are equally hard to calculate unless the channel possesses certain symmetries~\cite{pirandola2015ultimate}. The secret key rate is lower bounded by the two-way distillable entanglement. In order to decide whether the two-way key rate is nonzero it is thus sufficient to check whether $\om_{AB}$ is entangled (see~\cite{horodecki2009quantum}, Sections~XII.D, XII.F and XIV.A for an all-in-one overview). This can be decided via the PPT criterion which for $|A|=|B|=2$ is a sufficient and necessary condition. Currently, there also exists only a few upper bounds.  The upper bounds are harder to evaluate: to the authors' knowledge, none of them are tight and as we will see, they often require a non-trivial optimization step.

The first quantity we mention is the squashed entanglement introduced in~\cite{christandl2006quantum}
\begin{equation}\label{eq:SquashedEnt}
  {E_{\mathrm{sq}}(\om_{AB})}\df{1\over2}\inf_{\Scal}{I(A;B|E)_{\om}},
\end{equation}
where $\Scal_{E\to\tilde{E}}$ is a `squashing' channel and $I(A;B|E)$ is the conditional mutual information defined as
\begin{equation}\label{eq:SSA}
  0\leq I(A;B|E)\df H(AE)+H(BE)-H(ABE)-H(E).
\end{equation}
Squashed entanglement was adapted in~\cite{takeoka2014squashed} to define the squashed entanglement of a quantum channel~$\Ncal_{A'\to B}$
\begin{equation}\label{eq:SquashedEntChnl}
  E_{\mathrm{sq}}(\Ncal)\df\sup_{\vr_{AA'}}{E_{\mathrm{sq}}(\om_{AB})},
\end{equation}
where $\vr_{AA'}$ is the channel's input state (recall our convention of $\om_{AB}$ being the Choi matrix of $\Ncal$) and the following inequalities hold
\begin{equation}\label{eq:IneqChain2}
Q_\leftrightarrow(\Ncal)\leq K_\leftrightarrow(\Ncal)\leq E_{\mathrm{sq}}(\Ncal)\leq E_{\mathrm{F}}(\om_{AB}).
\end{equation}
The last inequality (cf.~\cite{christandl2006quantum}) is computationally the most relevant. The RHS is called the entanglement of formation (EOF) and was introduced in~\cite{bennett1996mixed}
\begin{equation}\label{eq:eof}
   E_{\mathrm{F}}(\om)\df\inf_{\om_{AB}}{\sum_i p_iH(A)_{\varpi_i}},
\end{equation}
where the optimization goes over the decomposition $\om_{AB}=\sum_{i}p_i\kbr{\varpi_i}{\varpi_i}_{AB}$. The EOF is normally difficult to calculate except in our case where a closed formula is known due to Wootters~\cite{wootters1998entanglement}.

Another notable bound is the regularized relative entropy of entanglement~\cite{vedral1997quantifying,audenaert2001asymptotic}
\begin{equation}\label{eq:REE}
  E_\mathrm{R}(\om)\df\lim_{n\to\infty}{1\over n}\min_{\xi_{A^nB^n}}{ D(\om_{A^nB^n}\|\xi_{A^nB^n})}
\end{equation}
where  $\xi_{A^nB^n}$ is a separable state and
\begin{equation}\label{eq:RelEntropy}
  D(\om\|\s)\df\Tr{}\big[\om\big(\log{\om}-\log{\s}\big)\big]
\end{equation}
is the quantum relative entropy~\cite{umegaki1962conditional}. Again, the regularization is an obstacle for its direct evaluation. Recently, the authors of~\cite{pirandola2015ultimate}, influenced by seminal~\cite{bennett1996mixed}, introduced the regularized relative entropy of entanglement
\begin{equation}\label{eq:REEofN}
  E_\mathrm{R}(\Ncal)\df\lim_{n\to\infty}{1\over n}\min_{\xi_{A^nB^n}}{ D(\om^{\otimes n}_{\Ncal}\|\xi_{A^nB^n})}
\end{equation}
of a channel $\Ncal_{A'\to B}$ where $\om_\Ncal$ is its Choi matrix. They showed that for a large class of channels (so-called `teleportation covariant') the following inequality holds
\begin{equation}\label{eq:REEstretch}
  E_\mathrm{R}(\Ncal)\leq\sup_{\Acal}\min_{\xi_{AB}} D(\om_{\Ncal}\|\xi_{AB}),
\end{equation}
where $\Acal$ are all adaptive strategies and $\xi_{AB}$ is a separable state. It turns out that the following inequalities hold:
\begin{equation}\label{eq:chain2}
Q_\leftrightarrow(\Ncal)\leq K_\leftrightarrow(\Ncal)\leq E_\mathrm{R}(\Ncal).
\end{equation}
Note that
\begin{equation}\label{eq:REvsEOF}
  \min_{\xi_{AB}} D(\om_{\Ncal}\|\xi_{AB})\leq E_{\mathrm{F}}(\om_{\Ncal})
\end{equation}
holds as well~(see~\cite{vedral1997quantifying}).

Finally, the authors of~\cite{christandl2016relative} defined the max-relative entropy of entanglement of a quantum channel $\Ncal_{A'\to B}$ as
\begin{equation}\label{eq:maxREEchnl}
  E_{\mathrm{max}}(\Ncal)=\sup_{\vr_{AA'}}{E_{\mathrm{max}}(\om_{\Ncal})},
\end{equation}
where $\vr_{AA'}$ is the channel's input state and in general we have
\begin{equation}\label{eq:maxREE}
  E_{\mathrm{max}}(\om_{AB})\df\min_{\xi_{AB}}{D_{\mathrm{max}}(\om_{AB}\|\xi_{AB})}.
\end{equation}
The states $\xi_{AB}$ are chosen from the set of separable states and the max-relative entropy $D_{\mathrm{max}}(\om\|\s)$ is  defined as~\cite{datta2009min}
\begin{equation}\label{eq:maxRE}
  D_{\mathrm{max}}(\om\|\s)\df\log{\|\s^{-1/2}\om\s^{-1/2}\|_\infty},
\end{equation}
where $\|M\|_\infty$ is the greatest eigenvalue of $M$. The max-relative entropy of entanglement was shown to satisfy
$$
Q_\leftrightarrow(\Ncal)\leq K_\leftrightarrow(\Ncal)\leq E_{\mathrm{max}}(\Ncal).
$$
Looking at the above definitions it does not come as a surprise that the upper bound could be difficult to find as well and so a simplified upper bound (on the upper bound)  was devised~\cite{christandl2016relative}:
\begin{equation}\label{eq:ChoiEB}
  E_{\mathrm{max}}(\Ncal)\leq B_{\mathrm{max}}(\Ncal)\df\min_{\xi_{AB}}{D_{\mathrm{max}}(\om_{\Ncal}\|\xi_{AB})},
\end{equation}
where $\xi_{AB}$ is a separable state representing a Choi matrix of an entanglement-breaking channel. Comparing (\ref{eq:maxREEchnl}) and~(\ref{eq:maxREE}) with~(\ref{eq:ChoiEB}) we see that we removed one optimization step that we would not be able to perform anyway as discussed earlier in this section.

\section{Results}\label{sec:results}

Ultimately, our goal is to maximize the secret key rate and distance for the physical parameters of the Unruh-DeWitt detectors such that they are reasonably realistic. This is a hard problem as the integrals~from Eqs.~(\ref{eq:J23forL}) must be solved numerically which itself is a rather time-consuming process if we want to make sure that the values we get can be trusted. Hence we choose the following strategy: for realistically chosen atomic parameters we made an educated guess and showcase our results by calculating the best rates we could find at the expense of a relatively small (but still respectable) distance and the best distance for a smaller but positive secret key rate. Especially here we got to an impressive distance in the order of thousands of kilometers so let's start with this case. We set the parameters introduced in Eqs.~(\ref{eq:UDWHint}) and~(\ref{eq:GaussianProfile}) to be $\s=5\times10^{-8}~\mathrm{s},a=1~\mathrm{GHz},\d=4\times 10^{8}~\mathrm{Hz}$ and $\la=0.363$.
\begin{figure}[h]
   \resizebox{13cm}{!}{\includegraphics{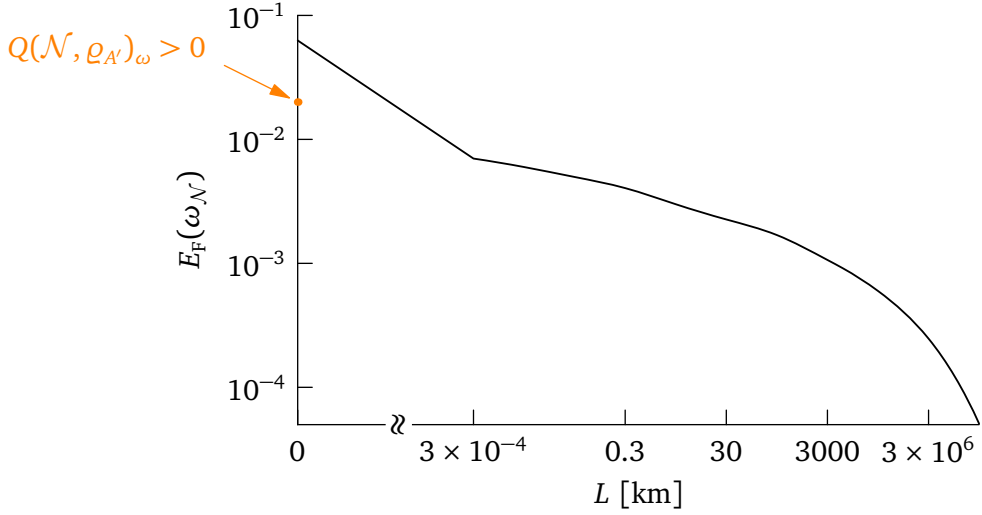}}
    \caption{The EOF given by $\s=5\times10^{-8}~\mathrm{s},a=1~\mathrm{GHz},\d=4\times 10^{8}~\mathrm{Hz}$ and $\la=0.363$ as an upper bound on the two-way quantum communication rate. The orange point denotes the one-way rate $Q(\Ncal,\varrho_{A'})_\omega\approx0.02$ for $L\gtrsim0$.}
    \label{fig:rateLADA}
\end{figure}
The EOF of the corresponding state $\om_{AB}$ is depicted in Fig.~\ref{fig:rateLADA}. We can see that it quickly decreases for $L>0$ but remains positive for more than a thousand kilometers indicating the presence of entanglement at large distances that can be used for two-way quantum communication. As discussed in the previous section, as long as $\om_{AB}$ is entangled the two-way secret key rate is strictly positive. But we do not know the actual rate of covert communication except with one truly remarkable exception. For the immediate vicinity of $L=0$ the one-way  quantum capacity is nonzero: $Q(\Ncal,\vr_{A'})_\om>0$ for $0\leq L\leq3\times10^{-29}~\mathrm{m}$.   This result is useless for covert communication for large distances but nonetheless conceptually important. $Q(\Ncal,\vr_{A'})_\om$ provides not only an achievable lower bound on the one-way secret key rate as indicated in Eq.~(\ref{eq:IneqChain1}) but mainly we can exploit $Q(\Ncal)=Q_{\varnothing}(\Ncal)$~\cite{bennett1996mixed,barnum2000quantum}. That is, the participants do not have to classically communicate at all. They have to initially agree on a decoding strategy without the need to talk again in order to quantum communicate. One could wonder why the channel is not an identity for $L\to0$. This is because $L=0$ does not mean a measurement at the same time and especially on the same system. It rather corresponds to the situation depicted in the left panel of Fig.~\ref{fig:timelike}. The information is sent to the future where it is received at the same spatial position but by a completely different observer (in his own lab that could have been built in the meantime).

\begin{figure}[h]
   \resizebox{15cm}{!}{\includegraphics{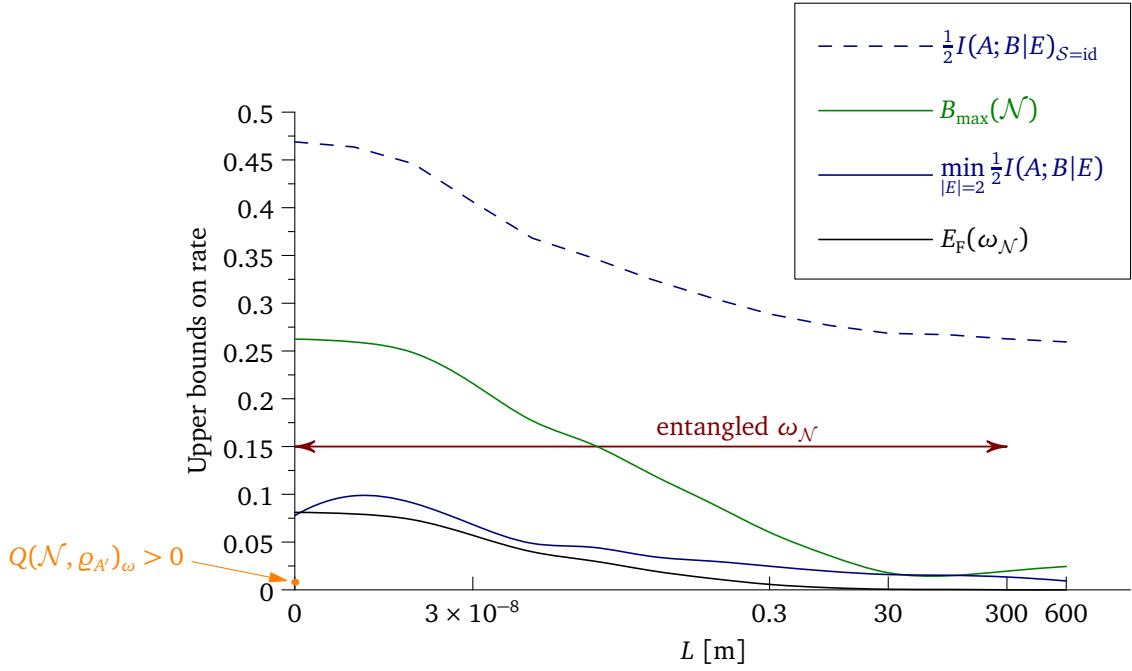}}
    \caption{Three different upper bounds on the two-way secret key rate. The blue curves are based on the squashed entanglement. The green curve is an upper bound on the max-relative entropy and finally the best result (black) is given by the entanglement of formation. The orange point denotes the one-way rate $Q(\Ncal,\varrho_{A'})_\omega\approx0.008$ for $L\gtrsim0$. The $x$ axis is scaled neither linearly nor logarithmically.}
    \label{fig:rateL}
\end{figure}

\begin{figure}[t]
   \resizebox{10cm}{!}{\includegraphics{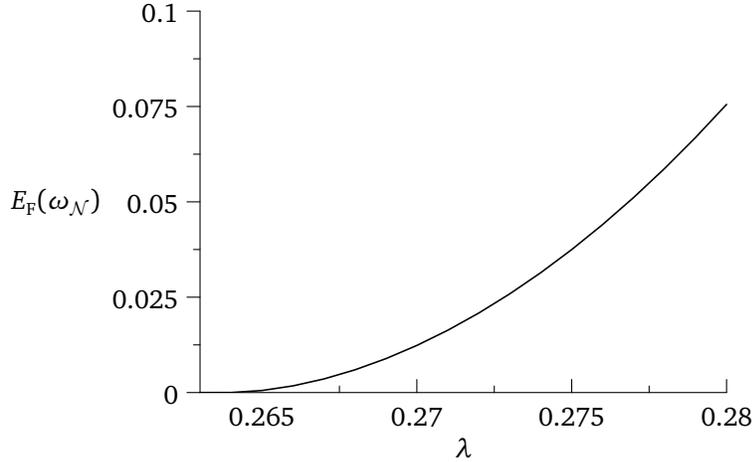}}
    \caption{The dependence of the EOF on the coupling constant $\la$ for $a\s=98,\s\d=30$, and $aL=0$. }
    \label{fig:rateLAM}
\end{figure}

By tuning the detectors' parameters we can increase the EOF (and most likely the rate as well). The EOF is plotted in Fig.~\ref{fig:rateL} for the following set of parameters: $\s=12\times10^{-8}~\mathrm{s},a=1/5\times10^9~\mathrm{Hz},\d=11/20\times 10^{8}~\mathrm{Hz}$ and $\la=0.581$. For $L=0$ we get a better result compared to Fig.~\ref{fig:rateLADA} but it drops to zero. This is the area marked by an arrow in Fig.~\ref{fig:rateL}. The zero point coincides with $\om_{\Ncal}$ becoming separable. As in the previous case, the dimension of the detectors' state $\om_{AB}$ makes it easy to find an upper bound via the EOF but it is also a good opportunity to test how tight bounds we would have gotten  if we could not rely on the calculable EOF.  First, we plotted the unoptimized squashed entanglement Eq.~(\ref{eq:SquashedEnt}) where  $\Scal=\id$ was set (see the blue dashed curve). It is clearly an inappropriate bound which is not surprising given the trivial squashing channel. To get a better upper bound we numerically optimized over the system $E$ such that $|E|=2$ using the tools from~\cite{opti}. This provides a better estimate and surprisingly, sometimes (for $L=0$) even better than the one given by the EOF.  Another plotted  estimate is the simplified upper bound on the max-relative entropy of entanglement Eq.~(\ref{eq:ChoiEB}) we calculated numerically (green curve). Overall, it  grossly overestimates the EOF but provides a better estimate when $\om_\Ncal$ is becoming separable. The third of the mentioned estimates, Eq.~(\ref{eq:REEstretch}), is not known to be applicable to our channel.
We can also study how the coupling strength $\la$ affects the rate. For the parameters $a\s=98,\s\d=30$, and $aL=0$ (we set $L=0$) we plot the EOF against $\la$ in~Fig.~\ref{fig:rateLAM}. As expected, with a decreasing coupling constant $\la$ the rate goes down as well until the state becomes so weakly coupled that it is separable and the rate is zero.

\subsection*{Eavesdropping}

We will comment on an interesting problem of what eavesdropping actually means in our situation. No quantum signal is exchanged between the communicating parties  and so an eavesdropper cannot take  direct advantage of the QKD paradigm of how the security is analyzed: Normally, any kind of channel noise is attributed to the eavesdropper and the value of the secret key rate reflects the fact whether under this assumption information-theoretic security can be achieved. In our opinion there are two possibilities of how Eve can interact with the system held by the legitimate parties and none of them affects the obtained results. Eve could gain access to the labs and tamper with the detectors. But this is really an out-of-the-scope attack, something more similar to a side-channel attack in QKD where an eavesdropper manipulates the detector's efficiency by sending laser pulses~\cite{lo2014secure}. A more relevant attack would be Eve setting up her own detectors at various spacetime points. In principle she could be in possession of a multipartite entangled state $\om_{ABE_1\dots E_n}$ or even a purification of the state shared by the sender and the receiver. But this does not help her at all since she cannot manipulate the marginal state $\om_{AB}$. As a matter of fact, this is precisely the assumption in the security of QKD -- an eavesdropper holds the purification~\cite{rennersecurity} (or the complementary output of the channel~\cite{devetak2005private}) and the secret key rate is calculated or estimated under these most advantageous conditions for the eavesdropper.

\section{Conclusions  and open problems}\label{sec:discussion}

We developed a truly fundamental covert quantum communication protocol by exploiting the properties
of Minkowski vacuum. Contrary to the previously introduced schemes, the covert communication does
not need to be hidden in the environmental (thermal) noise and is not limited to a particular model
of an optical quantum channel. What makes our results fundamental is that the rate of covert communication in the asymptotic limit of quantum
communication is in fact strictly positive; where previous results tended towards zero in the same limit. In our scheme no quantum signal is
exchanged at all which is possible due to intrinsically quantum properties of Minkowski vacuum essentially offering preshared entangled states at large spatial distances. We tapped into this resource by
equipping two inertial observers by specially crafted Unruh-DeWitt detectors (following~\cite{olson2012extraction}) and calculating
their response up to the fourth perturbative order. We found that the detectors are entangled and we
analyzed the meaning of this entanglement from the quantum communication point of view. We found that
for suitable parameters and using two-way quantum communication, that a positive secret key rates can be achieved over
thousands of kilometers.

The main open problem is whether we can avoid classical communication altogether for greater distances. We have shown that quantum information can be covertly sent to the same place in the future without the need to classically communicate whatsoever. This is possible because of the positive value of the coherent information for the corresponding quantum channel. To answer this question we would need to go well beyond the fourth perturbative order -- a strategy that would bring the answer to another open problem: What is the actual non-perturbative channel? If $\om_{AB}$ is interpreted as a Choi-Jamio\l kowski matrix as was advocated here, then its form strongly resembles a qubit Pauli channel. This hardly can be true for an arbitrary value of the detector parameters since in order for it to be a Pauli channel the first and last diagonal elements $\om_{AB}$ must be equal. But perhaps in some regime or limit this could indeed be the case. 

\section*{Acknowledgements}
The authors thank Nathan Walk and Timothy Ralph for comments.

\bibliographystyle{unsrt}


\end{document}